\documentclass[showpacs,superscriptaddress,amssymb,preprint]{revtex4}
\usepackage{amssymb}
\usepackage{amsmath}
\usepackage{mathrsfs}
\usepackage{graphicx}
\usepackage{txfonts}

\begin{document}

\draft

\title{Entanglement and Berry Phase in Two Interacting Qubits }

\author{Ai Min Chen  }
\affiliation{Centre for Modern Physics and Department of Physics,
Chongqing University, Chongqing 400044, The People's Republic of
China}
\author{Sam Young Cho}
\email{sycho@cqu.edu.cn}
 \affiliation{Centre for Modern Physics and
Department of Physics, Chongqing University, Chongqing 400044, The
People's Republic of China}
\author{Taeseung Choi}
\email{tschoi@swu.ac.kr}
\affiliation{Seoul Women's University,  Seoul,
139-774, Korea}

\date{\today}

\begin{abstract}
 Entanglement and Berry phase
 are investigated in two interacting qubit systems.
 The XXZ spin interaction model with a slowly rotating magnetic field
 is employed for the interaction between the two qubits.
 We show how the anisotropy of interaction reveals
 unique relations between the Berry phases and the entanglements
 for the eigenstates of the system.
\end{abstract}


\maketitle

 {\it Introduction.}
 Entanglement of a quantum system have been intensively studied
 in a wide range of research areas.
 Various systems have been considered to investigate
 a key role of entanglement in quantum phenomenon
 such as quantum phase transitions \cite{Osborne,Wu,Oliveira},
 many-body effects \cite{manybody,Cho_Kondo},
 quantum information processing \cite{computation,teleportation,cryptographic},
 and quantum transport \cite{Sim,Zhang,Loss}.
 During the periodic time evolution of the system, i.e.,
 the system Hamiltonian is varied slowly and eventually brought back
 to its initial form, the entanglement has been shown to affect
 on the geometric phase of the system.
 For entangled bipartite systems, especially,
 geometric phases have been studied \cite{Sjovist,Tong,Xiang}.
 It has been shown
 in bipartite systems that a prior entanglement influences on the geometric
 phase although there are no interactions during the cyclic evolution
 of the system.

 Recently, advanced quantum technologies have made it possible
 to manipulate a quantum system in a controllable manner.
 The interactions between the subsystems of a composite
 system can be controlled by varying the system parameters
 \cite{Cho07}.
 For example, superconducting qubits have been realized
 in experiments
 and various types of interactions between them have been
 demonstrated such as Ising-\cite{Pashkin,Izmalkov,Majer},
 XY-\cite{Niskanen,Berkley,Steffen}, and XXZ-type \cite{Shi} interactions
 for two qubit systems.
 In addition, a possible way to generate a Berry phase
 has been shown theoretically in flux qubits \cite{Peng}.
 Thus, in such quantum systems,
 the interaction strengths between the subsystems of a composite system
 may be adjusted by controlling system parameters.
 Varying the interaction strengths
 may make the entanglement of the subsystems changing \cite{KimCho,Kim07}.
 Also, phase dynamics of the subsystems can be changed
 by varying interaction strengths between the subsystems \cite{Yi}.

 Of particular interest are the Berry phases \cite{Berry} of a composite system
 because its applications are to be the implementation of quantum
 information processing \cite{Jones,Ekert,Falci,Wang}.
 This raises questions of how the interactions
 among the subsystems changes the Berry phase and entanglement of the
 composite system and what the relation is between the Berry phase
 of the composite system and the entanglement of the two subsystems.
 In this Brief Report,
 we consider a composite system consisting of two interacting
 qubits (spin-$\frac{1}{2}$'s). We focus on the effects
 of interactions
 on relations between the Berry phases and entanglements
 for the eigenstates of the system.
 To do this,
 the XXZ-type of spin exchange interaction is employed
 to describe the interaction between two qubits.
 We investigate the behavior of the Berry phase
 and entanglement of
 two interacting qubits
 due to a rotating magnetic field.
 The spin exchange interaction effects
 on the Berry phase and entanglement are discussed.
 A relation between the Berry phase and entanglement
 for the eigenstates of the systems is found to be unique as
 the interaction strengths vary.

 {\it Model.}
 We start with two interacting qubits corresponding to
 two spin-$\frac{1}{2}$'s, $\textbf{S}_{1}$ and $\textbf{S}_{2}$,
 in an external magnetic field. The system is described by
 the XXZ spin Hamiltonian,
 \begin{equation}
 H = H_x + H_z + H_B,
 \label{Hamiltonian}
 \end{equation}
  where $H_x=J_{x}(S^{x}_{1}S^{x}_{2}+S^{y}_{1}S^{y}_{2}) $,
  $H_z=J_{z}S^{z}_{1}S^{z}_{2}$, and
  $H_B=\mu(\textbf{S}_{1}+\textbf{S}_{2})\cdot\textbf{B}(t)$.
 Here, $\mu$ is the gyromagnetic ratio
 and $J_{x}$ and $J_{z}$ are the spin exchange interaction strengths
 between the spins.
 The slowly rotating magnetic field $\textbf{B}(t)=B\ \hat{\bf n} (t)$
 is chosen
 with the unit vector
 $\hat{\textbf{n}}(t)=(\sin\theta\cos\phi(t),\sin\theta\sin\phi(t),\cos\theta)$.
 It is assumed that the frequency of rotating magnetic field
 is a constant $\omega$ and then $\phi(t)=\omega t$.
 For a period of time $T$, the $\phi$ is slowly changing from $\phi(0)=0$
 to $\phi(T)=2\pi$.

 {\it Eigenstates of the system.}
 There are four eigenvalues $E_n$ ($n\in\{0,1,2,3\}$) of the
 Hamiltonian.
 One of them is unique, i.e., $E_0=-(J_z+2 J_x)$.
 This eigenstate is nothing but the spin singlet state,
$  |\Psi_0 \rangle=  \frac{1}{\sqrt{2}}
  \left(\left|\uparrow\downarrow \right\rangle
   - \left|\downarrow\uparrow \right\rangle\right) $.
 Note that the singlet state is not a function of the interaction
 parameters.
 Since the Bell states are maximally entangled states
 for two spin (qubit) systems,
 this singlet state is maximally entangled for all
 interaction parameter regimes.
 The singlet state does not capture any extra phase, i.e.,
 its Berry phase is zero, during the adiabatic and cyclic evolution
 of the system.

 For other eigenvalues, there is a cubic function having a form,
 \begin{equation}
 F(\varepsilon, J)= \varepsilon^3- 2 J\ \varepsilon^2
            -4B^2_0\ \varepsilon + 8 J B^2_0 \cos^2\!\theta,
 \label{cubic}
 \end{equation}
 where  $B_0=\mu B/2$.
  When $Q = 4(J^2 + 3 B^2_0)/9$
 and $R=4J(2J^2 +9 B_0^2(1 -3 \cos^2\theta))/27$
 are defined, for $R^2 \leq Q^3$,
 the three real roots of $F(\varepsilon,J)=0$ are given by
 \begin{equation}
 \varepsilon_n
  = 2 \sqrt{Q} \cos\left(\frac{P+2 n \pi}{3} \right)+
  \frac{2}{3}J,
 \end{equation}
 where  $P =\cos^{-1} \left( R/\sqrt{Q^3}\right)$.
 Then, the eigenvalues $E_n=\varepsilon_n +J_z$ of
 the Hamiltonian in Eq. \ref{Hamiltonian}
 satisfy
 \begin{equation}
 F(\varepsilon_n=E_n-J_z, J=J_x-J_z)=0.
 \end{equation}
 For the eigenvalues $E_n$ $(n \in \{ 1, 2, 3 \})$, generally,
 the other instantaneous eigenstates are
 in a superposition of the three triplet states,
\begin{subequations}
 \begin{equation}
 |\Psi_n\rangle
  =
        a_n e^{-i\phi} \left|\uparrow\uparrow\right\rangle
      + \frac{b_n }{\sqrt{2}}
       \Big( \left|\uparrow\downarrow\right\rangle
               +\left|\downarrow\uparrow\right\rangle \Big)
      +c_n e^{i\phi} \left|\downdownarrows\right\rangle, 
 \label{eigenv1}
 \end{equation}
 where the coefficients in terms of the energy $\varepsilon_n$ are given by
 \begin{eqnarray}
  a_n  &=& -\frac{2}{\sqrt{d_n}}
              B_0\sin\theta \left(\varepsilon_n +2B_0\cos\theta\right), \\
  b_n &=& -\sqrt{\frac{2}{ d_n}}
          \left(\varepsilon_n +  2B_0\cos\theta\right)
          \left(\varepsilon_n - 2B_0\cos\theta\right) , \\
  c_n  &=& -\frac{2}{\sqrt{d_n}}
                 B_0\sin\theta \left(\varepsilon_n
                 -2B_0\cos\theta\right)
 \end{eqnarray}
 with
 $
   d_n = 2(\varepsilon_n^4 -
       2(1+3\cos 2\theta)B_0^2 \ \varepsilon^2_n+  16B_0^4\cos^2\!\theta).
 $
\end{subequations}
 It should be noticed that the coefficients are a function of
 the shifted energies as $\varepsilon_n=E_n-J_z$.
 $\varepsilon_n$ is a function of the exchange energy difference
 between $J_x$ and $J_z$, i.e.,
 $\varepsilon_n(J)$, where $J=J_x-J_z$.
 Actually, the $\varepsilon_n$ determines the behaviors of the coefficients
 as the interactions vary.
 Then,
 the anisotropy of the exchange interactions plays
 a significant role for entanglement and Berry phase.

 Note that the coefficients have a property of
 $a_n(\theta) = c_n(\pi-\theta)$ and $b_n(\theta)=b_n(\pi-\theta)$.
 In addition, when the energy $\varepsilon_n$ changes to
 $-\varepsilon_n$, the coefficients
 hold the relations of $a_n(\varepsilon_n) = -c_n(-\varepsilon_n)$ and
 $b_n(\varepsilon_n)=b_n(-\varepsilon_n)$.
 As a consequence, the coefficients satisfy
 $a_n(\varepsilon_n,\theta) = -a_n(-\varepsilon_n, \pi-\theta)$
 and $b_n(\varepsilon_n,\theta)=b_n(-\varepsilon_n,\pi-\theta)$.

 \begin{figure}
 \vspace*{4.2cm}
 \includegraphics{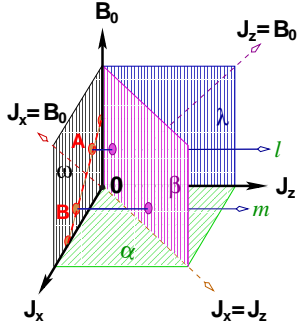}
 \caption{(Color online) Interaction parameter space
   for XXZ spin model with
   an external magnetic field. $J_x$ and $J_z$ are
   the strengths of spin exchange interactions.
   $B_0$ is the interaction energy between the spins and
   the applied magnetic field.
   There are four characteristic parameter planes
   for the two interacting spins.
   Note that, (i) for $B_0=0$, i.e., the $\alpha$-plane,
   the XXZ Hamiltonian without the magnetic field
   has the eigenstates which do not depend on the spin
   exchange interactions.
   (ii) Also, for the isotropic spin interaction
   $(J_x=J_z)$
   with the magnetic field $B_0 \neq 0$, i.e.,
   the $\beta$-plane, the eigenstates of the Heisenberg Hamiltonian
   do not depend on the strengths of the spin exchange interaction
   and the magnetic field.
   (iii) For $J_x=0$, i.e., the $\lambda$-plane,
   the system is described by the Ising Model with
   the magnetic field.
   (iv) On the other hand, for $J_z=0$, i.e., the $\omega$-plane,
   the XX model with the magnetic field describes the two interacting qubit
   systems.
   Controlling the interaction parameters will turn out a unique relation
   between entanglement and Berry phase for the eigenstates of the system
   based on the characteristic properties of the eigenstates
   in each interaction parameter plane.}
    \label{fig1}
 \end{figure}

 To help understanding of the model Hamiltonian
 in the interaction parameter space,
 a schematic diagram is drawn in Fig. \ref{fig1}.
 In the absence of magnetic field ${\bf B}(t)=0$, i.e.,
 the $\alpha$-plane in Fig. \ref{fig1},
 the system Hamiltonian reduces to the XXZ model without
 magnetic fields,
 $H_{\rm XXZ} =
 J_{x}(S^{x}_{1}S^{x}_{2}+S^{y}_{1}S^{y}_{2})+J_{z}S^{z}_{1}S^{z}_{2}$.
 Then, the eigenstates are to be a spin singlet state and
 three triplet states, i.e.,
  $|\Psi_n\rangle \in \left\{  \left|\uparrow\uparrow\right\rangle,
  \frac{1}{\sqrt{2}}
  \left(\left|\uparrow\downarrow \right\rangle
  \pm \left|\downarrow\uparrow \right\rangle\right),
  \left|\downarrow \downarrow \right\rangle
 \right\}$.
 The eigenstates of the XXZ model without magnetic
 fields do not depend on the interaction parameters $J_x$ and $J_z$.
 It is shown that, by varying the interactions between two
 qubits, the entanglements of the eigenstates
 for the XXZ spin exchange interaction without magnetic fields
 can not be manipulated.

 In the $\beta$-plane ($J_x=J_z$) of Fig. \ref{fig1},
 when magnetic fields ${\bf B}(t) \neq 0$ is applied and
 the spin exchange interaction is isotropic,
 the model Hamiltonian reduces to the Heisenberg model
 $H_{\rm H} =
 J_{H}(S^{x}_{1}S^{x}_{2}+S^{y}_{1}S^{y}_{2}+S^{z}_{1}S^{z}_{2})$
  with an external
 magnetic field $\textbf{B}(t)$.
 In this case, the energy $\varepsilon_n(J)$ is independent of
 the spin exchange interactions because
 $J=J_x-J_z=0$, i.e., $\varepsilon_n(0)$.
 Then the coefficients of the eigenstates are not a function of
 the spin exchange interactions.
 As a consequence, in terms of the bases
 $\left|\chi_+\right\rangle
  = \left(\begin{array}{c}
    e^{-i\phi}\cos \frac{\theta}{2} \\  \sin \frac{\theta}{2}
          \end{array} \right)$
          and
 $\left| \chi_-\right\rangle
  = \left(\begin{array}{c}
     -\sin \frac{\theta}{2} \\  e^{i\phi} \cos \frac{\theta}{2}
         \end{array}\right)$
           for a single spin,
 the eigenstates are given in the states
 $|\Psi_n\rangle \in \left\{
  \left|\chi_+\chi_+ \right\rangle,
  \frac{1}{\sqrt{2}}
 \left(\left|\chi_+\chi_- \right\rangle
              \pm \left|\chi_-\chi_+ \right\rangle\right),
 \left|\chi_-\chi_- \right\rangle \right\}$.
 They are nothing but a rotated one of the eigenstates
 of the XXZ Hamiltonian for ${\bf B} =0$.

 {\it Berry phase and concurrence.}$-$
 After the system undergoes an adiabatic and cyclic evolution
 with an initial state $|\Psi_n(0)\rangle$,
 the eigenstates have an additional phase, which comes from a
 geometrical feature, i.e., the Berry phase obtained by
 \begin{eqnarray}
 \gamma_n
  = \int_0^{2\pi}\!\!\! d\phi \left\langle\Psi_n\right| i\partial_\phi
            \left|\Psi_n\right\rangle
  = 2\pi\left(|a_n|^2-|c_n|^2\right).
 \label{berryp}
 \end{eqnarray}
 It is shown that the Berry phase is determined by the coefficients
 $a_n$ and $c_n$ of the eigenstates.
 From the relations of the coefficients, in general, the Berry phases
 satisfy $\gamma_n(\theta)=-\gamma_n(\pi-\theta)$
 and $\gamma_n(\varepsilon_n)=-\gamma_n(-\varepsilon_n)$.
 Also, the Berry phase holds a symmetry giving
 the relation $\gamma_n(\varepsilon_n,
 \theta)=\gamma_n(-\varepsilon_n,\pi-\theta)$.

  The entanglement of a pure general bipartite state
  can be quantified by introducing
  the concurrence \cite{concurrence}. For the eigenstates,
  the concurrences are given by
 \begin{eqnarray}
 C_n
 = \left|\langle \Psi_n|\sigma_y \otimes\sigma_y|\Psi^*_n\rangle\right|
 = \left|2 a_n c_n - b^2_n\right|,
 \end{eqnarray}
 where $\sigma_y$ is the pauli matrix.
 The concurrences $C_n$ range from 0 (an unentangled product state)
 to 1 (a maximally entangled state).
 Note that the entanglement of the eigenstates
 is not changed during the adiabatic and
 cyclic evolution of the system.
 The concurrences are a symmetry function with the axis
$\theta=\pi/2$,
 i.e., $C_n(\theta)=C_n(\pi-\theta)$.
 Changing the energy $\varepsilon_n$ to $-\varepsilon_n$
 gives the relation $C_n(\varepsilon_n)=C_n(-\varepsilon_n)$.
 In general, then, the entanglement of the eigenstates hold
 $C_n(\varepsilon_n,\theta)=C_n(-\varepsilon_n,\pi-\theta)$.

 {\it Relation between Berry phase and entanglement.}
 When the spin exchange interaction becomes very much anisotropic,
 i.e., (ii) $J_z \ll J_x, B_0$ for the $\omega$-plane
 and (iii) $J_x \ll J_z, B_0$ for the $\lambda$-plane in Fig. \ref{fig1},
 the XXZ model can be approximated to the XX and Ising models, respectively.

\begin{table}
\begin{tabular}{c|ccc}
  \hline\hline
  Interactions & $J_x \ll  B_0$ & $J_x \sim B_0$ & $ J_x \gg B_0$  \\
  \hline
  Hamiltonian & $H \simeq H_B$  & $H = H_x + H_B$ & $H \simeq H_x$  \\
  \hline
 $\left|\Psi_1\right\rangle$
      & $\left|\chi_-\chi_- \right\rangle$
      & $\cdots$
      & $\left|\downarrow\downarrow \right\rangle$
      \\
  $C_1$ & $0$ & $\cdots$ & $0$
    \\
  $\gamma_1[2\pi]$ & $-\cos\theta$ & $\cdots$ & $-1$
     \\ \hline
 $\left|\Psi_2\right\rangle$
      & $\frac{1}{\sqrt{2}}\left(\left|\chi_+\chi_- \right\rangle
                          +\left|\chi_-\chi_+ \right\rangle\right)$
      & $\cdots$
      & $-\left|\uparrow\uparrow \right\rangle$
           \\
  $C_2$ & $1$ & $\cdots$ & $0$
    \\
  $\gamma_2[2\pi]$ & $0$ & $\cdots$ & $1$
      \\ \hline
  $\left|\Psi_3\right\rangle$
      & $-\left|\chi_+\chi_+ \right\rangle$
      & $\cdots$
      & $-\frac{1}{\sqrt{2}}\left(\left|\uparrow\downarrow \right\rangle+\left|\downarrow\uparrow \right\rangle\right)$
        \\
  $C_3$ & $0$ & $\cdots$ & $1$
    \\
  $\gamma_3[2\pi]$ & $\cos\theta$ & $\cdots$ & $0$
     \\
  \hline\hline
\end{tabular}
\caption{ Comparison for eigenstates $|\Psi_n\rangle$
          ($n\in\{1,2,3\}$), their concurrences $C_n$ and Berry
          phases $\gamma_n$
          in varying interaction parameters $J_x$ and $B_0$
          of the XX model with an external magnetic field.
          The XX spin Hamiltonian with the magnetic field for the spins
          $\textbf{S}_{1}$ and $\textbf{S}_{2}$
           is
          $H = H_x + H_B$,
          where $H_x=J_{x}(S^{x}_{1}S^{x}_{2}+S^{y}_{1}S^{y}_{2}) $ and
          $H_B=\mu(\textbf{S}_{1}+\textbf{S}_{2})\cdot\textbf{B}(t)$.
          In the intermediate regime $ J_x \sim B_0$,
         `$\cdots$' indicates a superposition of the three triplet states.
}
 \label{table1}
\end{table}

 \begin{figure}
  \vspace*{3.5cm}
 \includegraphics{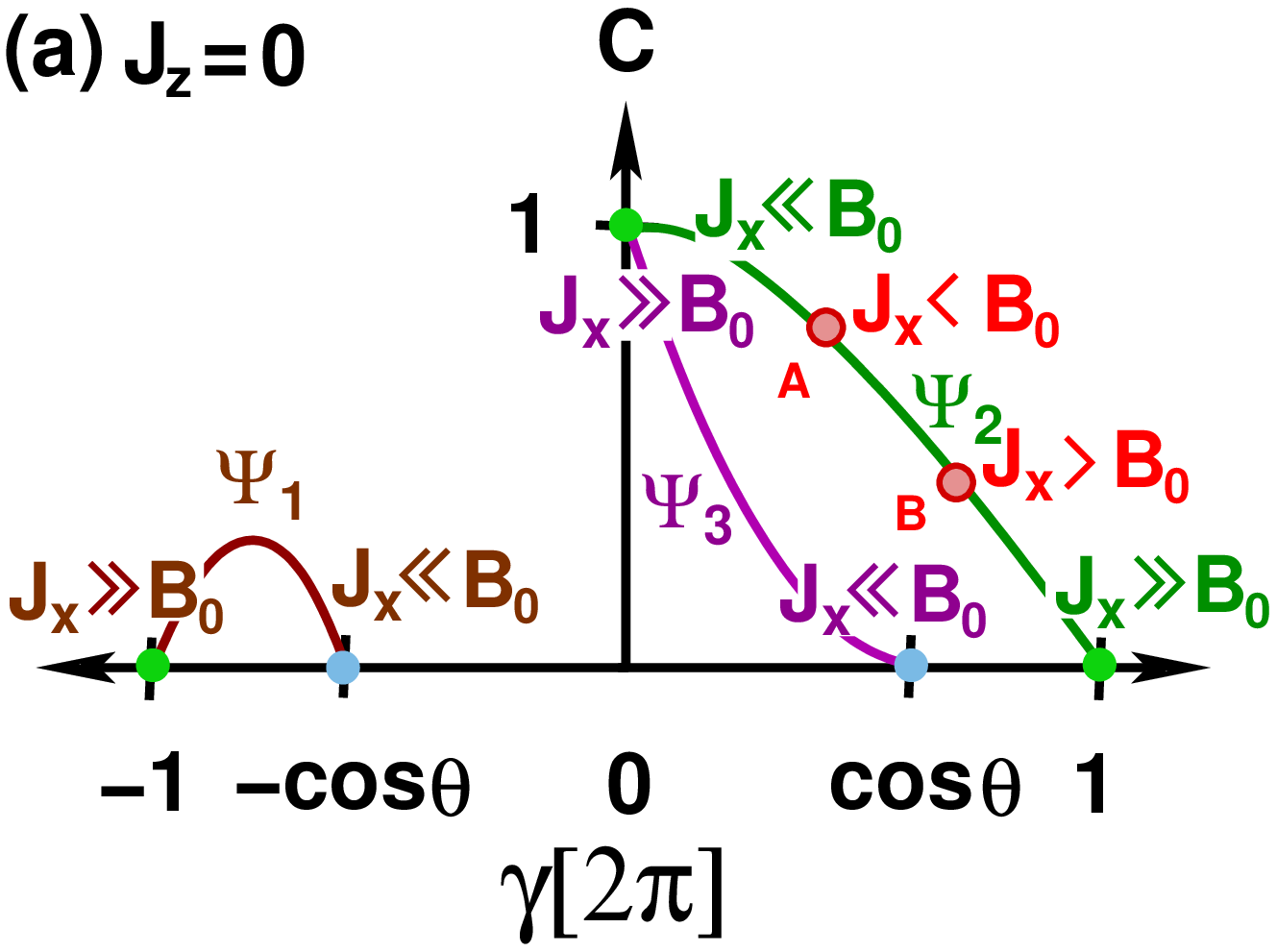}
 \includegraphics{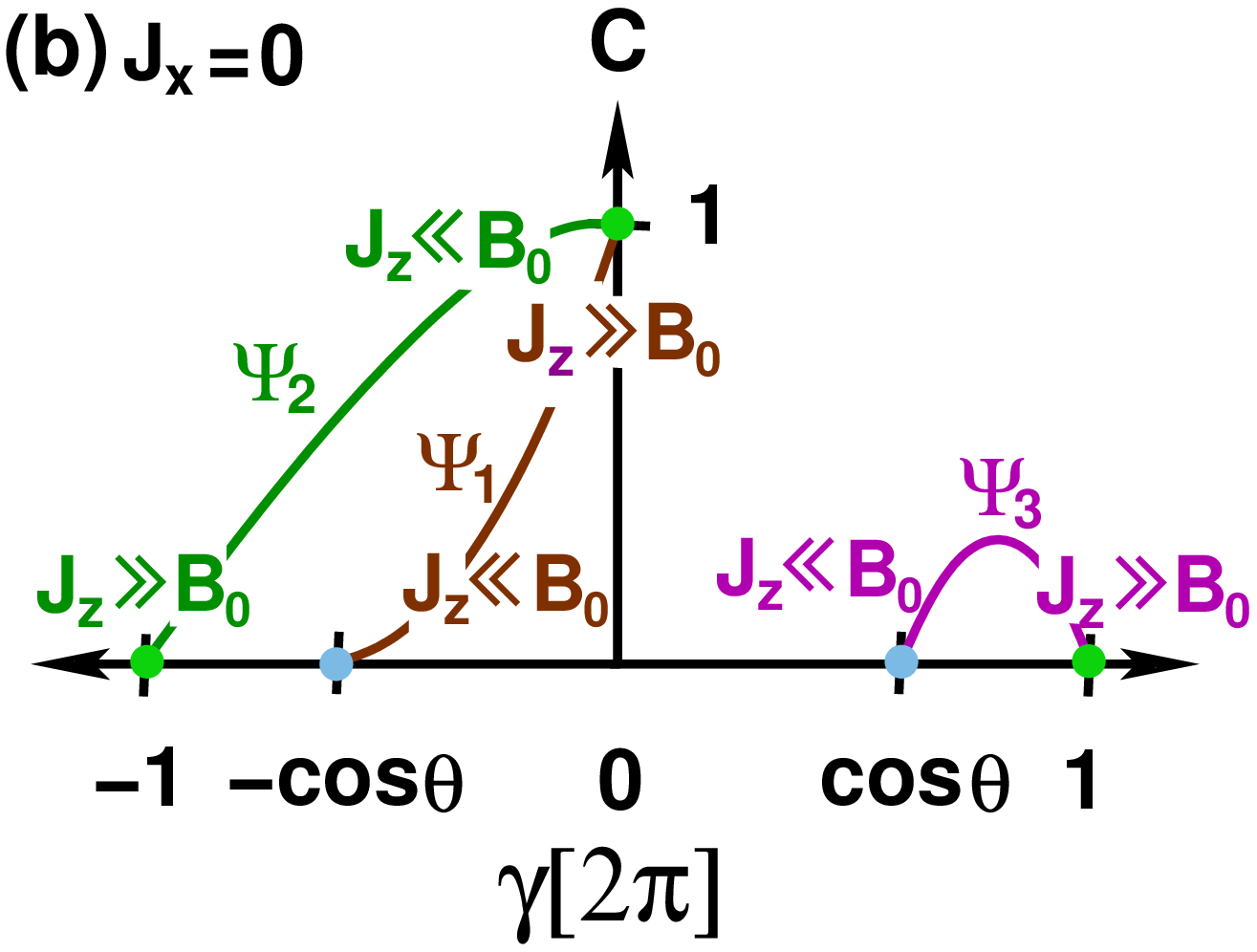}
%
 \caption{(Color online)
 Schematic diagrams for the relations between the concurrences $C_n$
 and Berry phases $\gamma_n$
 for the eigenstates $|\Psi_n\rangle$ ($n\in \{1,2,3\} $)
 of (a) the XX model $(J_z=0)$
 in varying the ratio $J_x$ to $B_0$
 and (b) the Ising model $(J_x=0)$
 in varying the ratio $J_z$ to $B_0$.}
  \label{fig2}
 \end{figure}

 Let us discuss the eigenstates for the XX model
 $H = H_x + H_B$, i.e., $J_z=0$ $(J > 0)$.
 The eigenenergies $E^{x}_n$ satisfy $F(\varepsilon^x_n=E^{x}_n, J_x)=0$.
 For $J_x \ll B_0$,
 the eigenstates become
 $|\Psi_1\rangle \simeq
  \left|\chi_-\chi_- \right\rangle
 $,
 $|\Psi_2\rangle \simeq
  \frac{1}{\sqrt{2}} \left(\left|\chi_+\chi_+ \right\rangle
              + \left|\chi_-\chi_+ \right\rangle\right)$, and
 $|\Psi_3\rangle \simeq -\left|\chi_+\chi_+ \right\rangle$.
 In the limit of $J_x \gg B_0$, the eigenstates are to be
 $|\Psi_1\rangle \simeq \left|\downarrow \downarrow \right\rangle$,
 $|\Psi_2\rangle \simeq -\left|\uparrow\uparrow\right\rangle$, and
 $|\Psi_3\rangle \simeq
  -\frac{1}{\sqrt{2}}
  \left(\left|\uparrow\downarrow \right\rangle
  + \left|\downarrow\uparrow \right\rangle\right)$.
 Table \ref{table1} shows the relation between the Berry phase and
 concurrence for the limiting values.
 The coefficients of the eigen wavefunctions are a monotonic function
 of $\varepsilon_n$ and the interactions.
 In Fig. \ref{fig2} (a), the overall behaviors of the relations
 are schematically drawn as the ratio of $J_x$ to $B_0$ varies
 for the eigenstates once the polar angle $\theta$ is fixed.
  As $J_x$ varies from $B_0 \gg J_x$,
 $|\Psi_2\rangle$ ($|\Psi_3\rangle$)
 reaches gradually to a product state (maximally entangled state)
 from a maximally entangled state (product state),
 while  $|\Psi_1\rangle$ changes from
 a product state to another product state.
 During the adiabatic and cyclic evolutions,
 $|\Psi_2\rangle$ ($|\Psi_3\rangle$)
 captures a Berry phase up to zero ($2\pi$)
 from $\cos\theta$ (zero)  as $J_x$ varies from $B_0 \gg J_x$.
 while  $|\Psi_1\rangle$ takes a Berry phase from $-\cos\theta$ to $-2\pi$.
 $|\Psi_1\rangle$ does not reach any maximally entangled state.
 It is shown that
 the Berry phase is zero when an eigenstate become a maximally entangled
 state.
 It was confirmed numerically.

\begin{table}
\begin{tabular}{c|ccc}
  \hline\hline
  Interactions & $J_z \ll  B_0$ & $J_z \sim B_0$ & $ J_z \gg B_0$  \\
  \hline
  Hamiltonian & $H \simeq H_B$  & $H = H_z + H_B$ & $H \simeq H_z$  \\
  \hline
 $\left|\Psi_1\right\rangle$
      & $\left|\chi_-\chi_- \right\rangle$
      & $\cdots$
      & $-\frac{1}{\sqrt{2}}\left(\left|\uparrow\downarrow \right\rangle+\left|\downarrow\uparrow \right\rangle\right)$
      \\
  $C_1$ & $0$ & $\cdots$ & $1$
    \\
  $\gamma_1[2\pi]$ & $-\cos\theta$ & $\cdots$ & $0$
     \\ \hline
 $\left|\Psi_2\right\rangle$
      & $\frac{1}{\sqrt{2}}\left(\left|\chi_+\chi_- \right\rangle
                          +\left|\chi_-\chi_+ \right\rangle\right)$
      & $\cdots$
      & $\left|\downarrow\downarrow \right\rangle$
           \\
  $C_2$ & $1$ & $\cdots$ & $0$
    \\
  $\gamma_2[2\pi]$ & $0$ & $\cdots$ & $-1$
      \\ \hline
  $\left|\Psi_3\right\rangle$
      & $-\left|\chi_+\chi_+ \right\rangle$
      & $\cdots$
      & $-\left|\uparrow\uparrow \right\rangle$
        \\
  $C_3$ & $0$ & $\cdots$ & $0$
    \\
  $\gamma_3[2\pi]$ & $\cos\theta$ & $\cdots$ & $1$
     \\
  \hline\hline
\end{tabular}
\caption{ Comparison for eigenstates $|\Psi_n\rangle$
          ($n\in\{1,2,3\}$), their concurrences $C_n$
          and Berry phases $\gamma_n$
          in varying interaction parameters $J_x$ and $B_0$
          of the Ising model with an external magnetic field.
          The Ising spin Hamiltonian with the magnetic field for the spins
          $\textbf{S}_{1}$ and $\textbf{S}_{2}$
           is
          $H = H_z + H_B$,
          where $H_z=J_{z}S^{z}_{1}S^{z}_{2}$ and
          $H_B=\mu(\textbf{S}_{1}+\textbf{S}_{2})\cdot\textbf{B}(t)$.
          In the intermediate regime $ J_z \sim B_0$,
         `$\cdots$' indicates a superposition of the three triplet states.
}
 \label{table2}
\end{table}

 For $J_x=0$ $(J < 0)$, the XXZ model reduces to the Ising model,
 $H = H_z + H_B$.
 The eigenenergies $E^{z}_n$ satisfy $F(\varepsilon^z_n=E^{z}_n-J_z, -J_z)=0$.
  For $J_z \ll B_0$,
 the eigenstates become
 $|\Psi_1\rangle \simeq
  \left|\chi_-\chi_- \right\rangle
 $,
 $|\Psi_2\rangle \simeq
  \frac{1}{\sqrt{2}} \left(\left|\chi_+\chi_+ \right\rangle
              + \left|\chi_-\chi_+ \right\rangle\right)$, and
 $|\Psi_3\rangle \simeq -\left|\chi_+\chi_+ \right\rangle$.
 In the limit of $J_z \gg B_0$, the eigenstates are to be
 $|\Psi_1\rangle \simeq   -\frac{1}{\sqrt{2}}
  \left(\left|\uparrow\downarrow \right\rangle
  + \left|\downarrow\uparrow \right\rangle\right)$,
 $|\Psi_2\rangle \simeq \left|\downarrow \downarrow \right\rangle$, and
 $|\Psi_3\rangle \simeq -\left|\uparrow\uparrow\right\rangle$.
 Table \ref{table2} shows the relation between the Berry phase and
 concurrence for the limiting values.
 Since the cubic function in Eq. \ref{cubic}
 has a property $F(\varepsilon_n, J)=-F(-\varepsilon_n, -J)$,
 the coefficients of the eigenstates of the Ising model
 can be written in terms of the coefficients of the eigenstates of the XX model
 when $J_x$ is replaced with $-J_z$ and
 $\varepsilon^x_{1(3)}$ and $\varepsilon^x_{2}$
 are replaced with $-\varepsilon^z_{3(1)}$ and $-\varepsilon^x_{2}$,
 respectively.
 One finds the relations between the Berry phases for the XX and Ising models,
 $\gamma_{1(3)}^{x}(\varepsilon^x_{1(3)})
 =-\gamma_{3(1)}^{z}(\varepsilon^z_{3(1)})$,
 and $\gamma_{2}^{x}(\varepsilon^x_2)=-\gamma_{2}^{z}(\varepsilon^z_2)$.
 Also the relations between the concurrences for the XX and Ising
 models are given by
 $C_{1(3)}^{x}(\varepsilon^x_{1(3)})=C_{3(1)}^{z}(\varepsilon^z_{3(1)})$
 and
 $C_{2}^{x}(\varepsilon^x_2)=C_{2}^{z}(\varepsilon^z_2)$.
 Such a symmetrical property of the Berry phases and concurrences
 is shown by directly comparing Fig. \ref{fig2} (a)
 for the XX model and (b) for the Ising model.

 \begin{widetext}
\begin{table}
\begin{tabular}{c|cccccc}
  \hline\hline
  Interactions & $J_z \sim B_0 \ll J_x$ &
           $B_0 \ll J_z < J_x$& $B_0 \ll J_z = J_x$ & $B_0 \ll J_x < J_z$ & $B_0 \ll J_x \ll J_z$ \\
 \hline
  Hamiltonian & $H \simeq H_x$  & $\cdots$  &
              $H \simeq H_H$ & $\cdots$ & $ H \simeq H_z$  \\
 \hline
  $\left|\Psi_1\right\rangle$
      & $\left|\downarrow\downarrow \right\rangle$
      & $\cdots$
      & $\left|\chi_-\chi_- \right\rangle$
      & $\cdots$
      & $-\frac{1}{\sqrt{2}}\left(\left|\uparrow\downarrow \right\rangle+\left|\downarrow\uparrow \right\rangle\right)$ \\
   $C_1$
      & $0$
      & $\cdots$
      & $0$
      & $\cdots$ &  $1$ \\
   $\gamma_1[2\pi]$
      & $-1$
      & $\cdots$
      & $-\cos\theta$
      & $\cdots$  &  $0$ \\
 \hline
  $\left|\Psi_2\right\rangle$
      & $-\left|\uparrow\uparrow \right\rangle$
      & $\cdots$
      & $\frac{1}{\sqrt{2}}\left(\left|\chi_+\chi_- \right\rangle
                          +\left|\chi_-\chi_+ \right\rangle\right)$
      & $\cdots$
      & $\left|\downarrow\downarrow \right\rangle$ \\
   $C_2$
      & $0$
      & $\cdots$
      & $1$
      & $\cdots$ &  $0$ \\
   $\gamma_2[2\pi]$
      & $1$
      & $\cdots$
      & $0$
      & $\cdots$  &  $-1$ \\
 \hline
  $\left|\Psi_3\right\rangle$
      & $-\frac{1}{\sqrt{2}}\left(\left|\uparrow\downarrow \right\rangle+\left|\downarrow\uparrow \right\rangle\right)$
      & $\cdots$
      & $-\left|\chi_+\chi_+ \right\rangle$
      & $\cdots$
      & $-\left|\uparrow\uparrow \right\rangle$ \\
   $C_3$
      & $1$
      & $\cdots$
      & $0$
      & $\cdots$ &  $0$ \\
   $\gamma_3[2\pi]$
      & $0$
      & $\cdots$
      & $\cos\theta$
      & $\cdots$  &  $1$ \\
  \hline\hline
\end{tabular}
\caption{ Comparison for eigenstates $|\Psi_n\rangle$
          ($n\in\{1,2,3\}$), their concurrences $C_n$
          and Berry phases $\gamma_n$
          in varying interaction parameters $J_z$ for $B_0 \ll J_x$.
          The XXZ Hamiltonian with the magnetic field for the spins
          $\textbf{S}_{1}$ and $\textbf{S}_{2}$
           is
          $H = J_{x}(S^{x}_{1}S^{x}_{2}
          +S^{y}_{1}S^{y}_{2})+J_{z}S^{z}_{1}S^{z}_{2}
          + \mu(\textbf{S}_{1}+\textbf{S}_{2})\cdot\textbf{B}(t)$.
         (i) For $J_z \ll J_x$, the two spin Hamiltonian
         can be approximated to the XX model
         $H \simeq J_{x}(S^{x}_{1}S^{x}_{2}+S^{y}_{1}S^{y}_{2})$.
         (ii) For $J_z = J_x$, the Heisenberg model
         $H \simeq J(S^{x}_{1}S^{x}_{2}
          +S^{y}_{1}S^{y}_{2} + S^{z}_{1}S^{z}_{2})$ can describe
          the system.
         (iii) For $J_z \gg J_x$, the two spin Hamiltonian becomes the Ising
         model
         $H \simeq J_{z}S^{z}_{1}S^{z}_{2}$.
         In the intermediate regimes,
         `$\cdots$' indicates a superposition of the three triplet states.
}
 \label{table3}
\end{table}
 \end{widetext}

 \begin{figure}
  \vspace*{3.5cm}
 \includegraphics{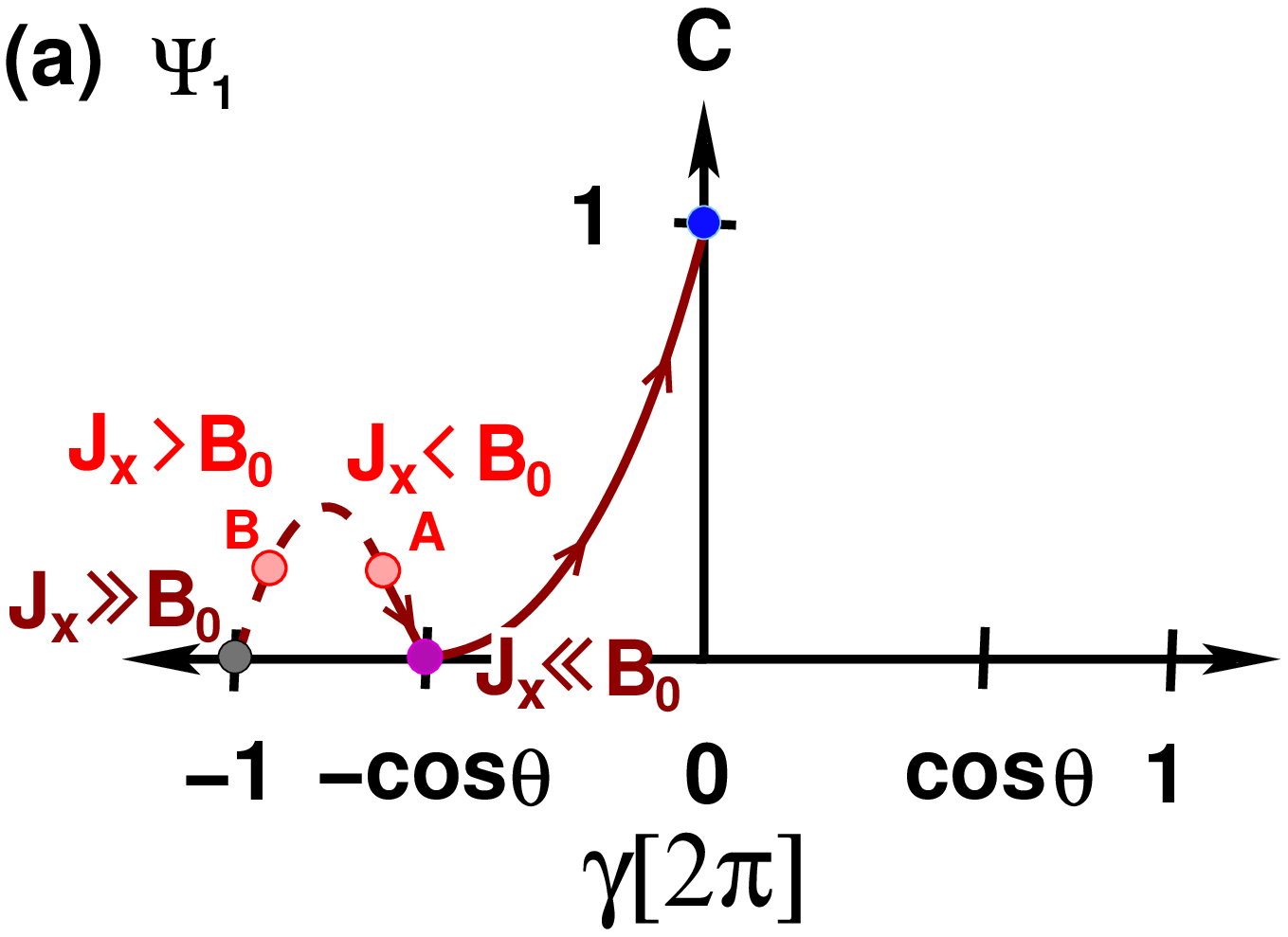}
   \vspace*{3.5cm}
 \includegraphics{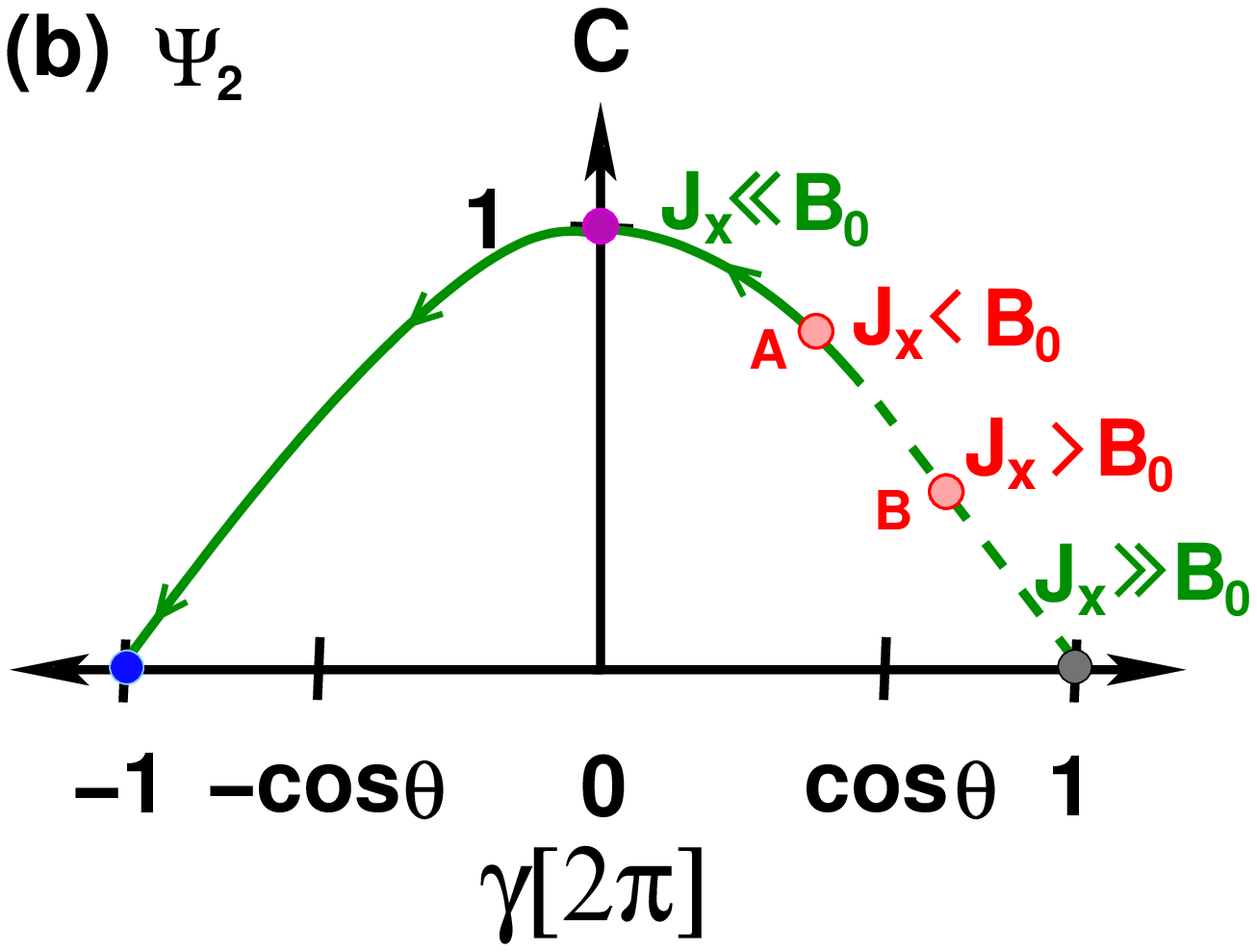}
 \includegraphics{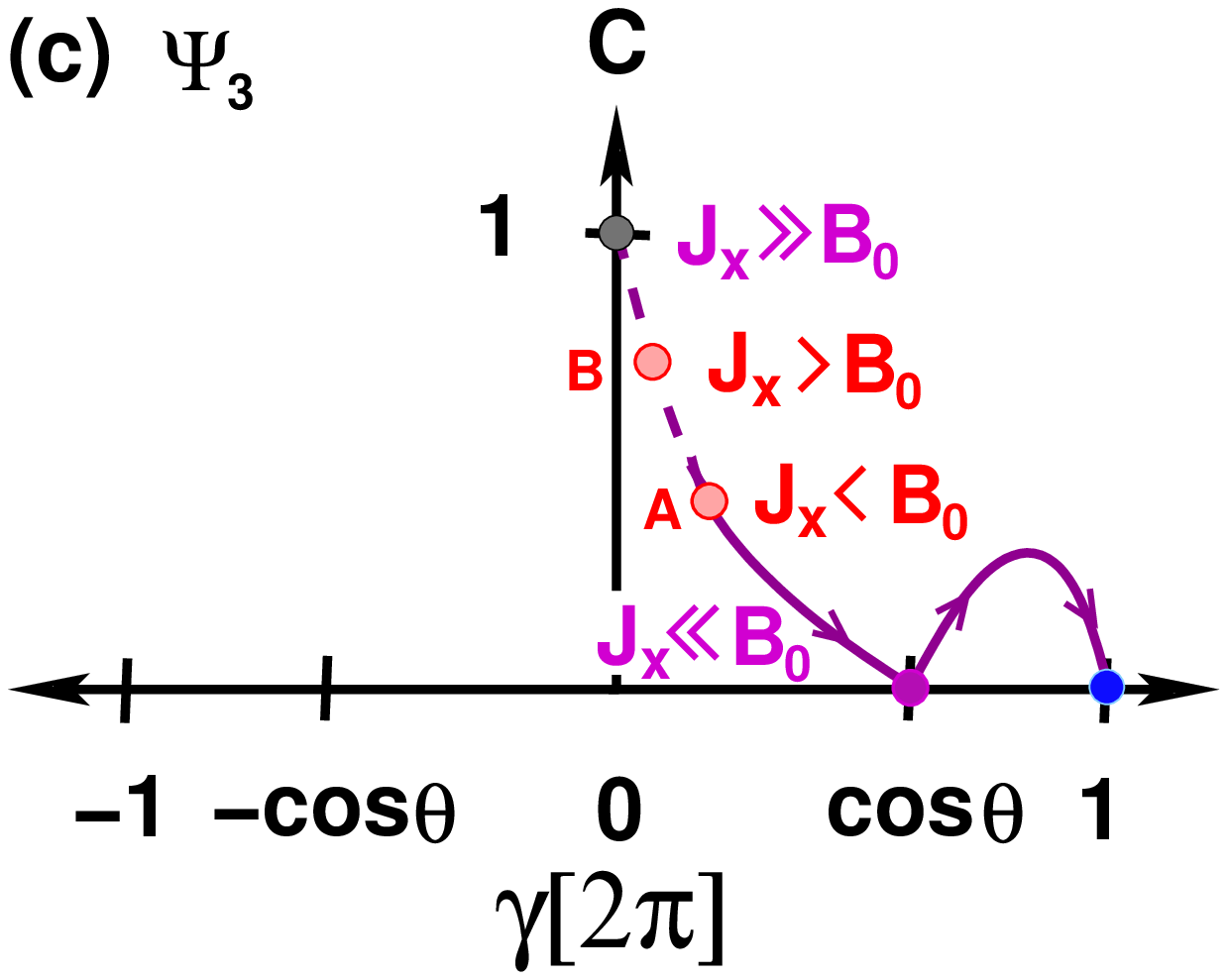}
%
 \caption{(Color online) Relations between Berry phases
  $\gamma_n$ and concurrences $C_n$
  of the eigenstates (a) $|\Psi_1\rangle$,
  (b) $|\Psi_2\rangle$, and (c) $| \Psi_3\rangle$
  in varying $J_z$ from zero for the XXZ model.
  The arrows on the curves indicate the increasing direction of
  $J_z$ from zero.
  When $J_z=0$, the starting point of the relation between the
  Berry phase and concurrence depends on the ratio $J_x$ to $B_0$.
  In Fig. \ref{fig1}, the $l$ ($J_x < B_0$) and $m$ ($J_x > B_0$)
  show the direction of increasing $J_z$ from zero.
  For instance,
  $\sf A$ ($J_x < B_0$) and $\sf B$ ($J_x > B_0$)
   shows that
  the starting points at $J_z=0$ is determined by the ratio $J_x$ to $B_0$.
  As $J_z$ varies from zero to $J_z=J_x$,
  the relation curve is the same with the case of XX model
  in Fig. \ref{fig2} (a).
  For $J_z > J_x$, the relation curve is the same with the case of
  Ising model in Fig. \ref{fig2} (b).
 }
 \label{fig3}
 \end{figure}

 Now, let us discuss more general cases.
 When the spin exchange interaction is isotropic, i.e,
 Heisenberg model ($J_x=J_z$), or the spin exchange interaction is very much
 anisotropic, i.e., XX ($J_x \gg J_z$) and Ising ($J_x \ll J_z$) models,
 as mentioned,
 the eigenstates are not
 dependent of the interaction parameters $J_x$ and $J_z$.
 Then, as $J_z$ varies, for $B_0 \ll J_x$,
 the behavior of the relation between the Berry phase and
 concurrence is summarized in the three characteristic limits
 in Table \ref{table3}.
 In Fig. \ref{fig3}, the schematic diagrams are then drawn
 the relations of the Berry phases and concurrences of the eigenstates
 as the exchange interaction $J_z$ varies from zero.
 The arrows on the curves of the relations
 indicate the direction of increasing $J_z$ from zero.
 At $J_z=0$, the values of the Berry phase and concurrence
 are on the relation for the case of XX model in Fig. \ref{fig2}
 (a). For instance,
 the {\sf A} ($J_x < B_0$) and {\sf B}($J_x > B_0$)
 are shown for $|\Psi_1\rangle$
 in Fig. \ref{fig2} (a). The values at these points {\sf A}
 and {\sf B} are the same in Fig. \ref{fig3} (b).
 The reason is why the energy $\varepsilon_n(J)$
 determining the Berry phases and concurrences
 is a function of the exchange energy difference $J=J_x-J_z$.
 In other words, once the polar angle $\theta$ is fixed,
 the relation between Berry phase and concurrence is uniquely
 determined by only one curve
 whatever we choose the values of the exchange energies
 and the magnetic energy $B_0$.
 Then if $B_0 \gg J_x$, the system corresponds to the Ising model
 in Fig. \ref{fig2} (b).
 As a result, we find that (i) for $J_x > J_z$,
 the behavior of the relation between the Berry phases and
 concurrences is similar to the XX model,
 (ii) while the behavior is similar to the Ising model for $J_x < J_z$.
 Therefore, the anisotropy of the exchange interaction determines
 the behavior of the relation between the Berry phase and
 entanglement in two interacting qubits (spins).

 {\it Summary.}
 The interaction effects on the entanglement and Berry phase
 are investigated in two qubits (spins). It is found that
 the anisotropy of the interaction plays an important role
 in determining the unique relation between the Berry phase and
 concurrence for the eigenstates of two interacting qubits.
 Also, it is shown that
 when the eigenstates become a maximally entangled state
 their Berry phases are zero.
 During the period time evolution of the system,
 unentangled eigenstates, as product states, can capture
 a finite value of Berry phase.

{\it Acknowledgments.}
 We thank Huan-Qiang Zhou and John Paul Barjaktarevic for helpful
 discussions.
 This work was supported by the National Science Foundation
 Project of CQ CSTC.
 This work was supported by a special research grant from Seoul Women's University (2008).

\end{document}